\documentclass[amsmath,amssymb,aps,pra,cite]{revtex4}
\usepackage{color}
 \allowdisplaybreaks

\begin{document}
%\twocolumngrid
\title{Orthogonal splitting in degenerate higher-order scalar-tensor theories}

\author{Z. Yousaf}
\email{zeeshan.math@pu.edu.pk}
\affiliation{Department of Mathematics, University of the Punjab,
Quaid-i-Azam Campus, Lahore-54590, Pakistan}

\author{M. Z. Bhatti}
\email{mzaeem.math@pu.edu.pk}
\affiliation{Department of Mathematics, University of the Punjab,
Quaid-i-Azam Campus, Lahore-54590, Pakistan}

\author{H. Asad}
\email{hamnaasad2596@gmail.com}
\affiliation{Department of Mathematics, University of the Punjab,
Quaid-i-Azam Campus, Lahore-54590, Pakistan}

\author{Yuki Hashimoto}
\email{s2471002@ipc.fukushima-u.ac.jp}
\affiliation{Faculty of Symbiotic Systems Science,
Fukushima University, Fukushima 960-1296, Japan}

\author{Kazuharu Bamba }
\email{bamba@sss.fukushima-u.ac.jp}
\altaffiliation{Corresponding author}
\affiliation{Faculty of Symbiotic Systems Science,
Fukushima University, Fukushima 960-1296, Japan}

%\date{}

\begin{abstract}
We explore a comprehensive analysis of the formalism governing the
gravitational field equations in degenerate higher-order
scalar-tensor theories. The propagation of these theories in the
vacuum has a maximum of three degrees of freedom and is at most
quadratic in the second derivative of the scalar field. We
investigate the gravitational field equation for spherically
symmetric anisotropic matter content along with its non-conserved
equations. Our analysis focuses on the evaluation of structure
scalars to assess their behavior under Einstein's modification. We
present a realistic mass contribution that sheds light on both
geometric mass and total energy budget evaluations for celestial
objects. Ultimately, we discuss two viable models restricted as
minimal complexity and conformal flatness to enhance the scientific
contribution of the present manuscript.
\end{abstract}
\maketitle

\section{Introduction}
The theory of General Relativity (GR)
\cite{einstein1915feldgleichungen,bonnor1981junction,scientific2016tests}
provides an all-encompassing explanation for the nature of space,
gravity and matter. The basic principles of the structure and
geometry of spacetime are covered by GR. The Friedmann equations,
derived from the Einstein field equations, depict how the universe
evolves in a homogeneous and isotropic spacetime. The conventional
model of the early universe, which involves epochs dominated by
radiation \cite{giudice2001largest} and matter
\cite{freese2002cardassian,kokubu2018effect}, can be adequately
explained through GR.

Modern astrophysics has brought to our attention the undeniable fact
that a significant amount of ``dark matter'' and ``dark energy''
\cite{huterer1999prospects,padmanabhan2003cosmological,amendola2010dark}
exist. These mysterious entities are necessary for justifying
observable cosmic dynamics
\cite{farajollahi2011cosmic,sami2012cosmological} and large-scale
structures
\cite{tegmark2004cosmological,springel2006large,carrasco2012effective}.
Without them, we may continually face a gap between theories and
observations. Therefore, acknowledging this discovery leads us
towards understanding the fundamental principles of the cosmos in an
entirely different level.

New approaches for classifying increasingly complex alternative
theories to Einstein's are needed for better analysis. With the
inclusion of extra degrees of freedom, modified Einstein models have
become harder to analyze. Therefore, it has become essential to
develop new techniques that can classify these diverse models and
make them more manageable. This can also help us to grasp the
attributes of the theory, problems that might arise, and its
observable effects. In short, it may enhance our insight into the
prevailing alternative theories of gravity
\cite{capozziello2008extended,capozziello2011extended,astashenok2015extreme}.

For creating models of inflation and dark energy, a scalar tensor
theories (STT) \cite{bronnikov1973scalar,goenner2012some} of gravity
could be considered helpful. These theories have been widely used in
attempts to get beyond GR because of their simplicity. Horndeski
\cite{rabochaya2016note,ogawa2020relativistic} developed STT by
combining Lagrangian with a scalar field. This scalar field involves
a second-order derivative. Also, it has the constraint that the
Euler-Lagrange equations cannot exceed second order. These are
four-dimensional theories that describe a scalar interacting with a
spin-2 tensor field. According to these theories, the metric
interacts with an additional scalar degree of freedom, which could
promote cosmic expansion. Scalar fields combined with gravitational
fields make up non-minimally coupled and higher-order components
through conformal transformations
\cite{maeda1989towards,capozziello1998recovering,gottlober1990sixth,crisostomi2016extended}.

Regarding velocities, the Hessian matrix of the Lagrangian is
referred to as ``degenerate." A degenerate Hessian matrix suggests
that there are main limitations since the system of momenta cannot
be reversed. The ``Degenerate Higher Order Scalar Tensor Theories''
(DHOST) were presented by Langlois and Noui
\cite{langlois2016degenerate}. Second-order derivatives of the
scalar field are permitted by DHOST theories. These theories permit
higher-order Euler-Lagrange equations, and they must only have one
scalar degree of freedom in order to prevent Ostrogradski
instabilities. These instabilities are frequently observed in
systems with higher-order time derivatives. Furthermore, they are
connected with additional degrees of freedom (ADF).

Crisostomi \emph{et al. } \cite{crisostomi2019cosmological} examined
cosmic solutions and the factors that affect their stability in
DHOST theory. Moreover, they investigated the fixed points of
dynamics and identified the prerequisites for the presence of a
de-Sitter attractor in late times. In modified gravity models
defined by DHOST theories, Boumaza \emph{et al.}
\cite{boumaza2020late} investigated the late cosmic development from
the non-relativistic matter-dominated period to the dark energy era.
Additionally, they determined the areas of parameter space where the
models exhibit perturbation stability. Thipaksorn and Karwan
\cite{thipaksorn2022cosmic} addressed the cosmic evolution using the
fixed points from the dynamical analysis of the scaling DHOST
theory. They determined that these theories also meet the condition
that the gravitational wave propagation speed is equal to the speed
of light. Furthermore, gravitational waves (GW) do not decay to dark
energy disturbances in addition to the scaling solutions.

Langlois \cite{langlois2019dark} considered the self-gravitating
objects in the context of DHOST theories. Moreover, he observed that
these theories could be an efficient approach to explain the dark
energy in cosmology. Achour \emph{et al.} \cite{achour2020hairy}
analyzed the prospect of exploring new areas of the solution space
of DHOST theories through the implementation of disformal field
redefinitions. Furthermore, they evaluated various novel solutions.
These solutions include asymptotically locally flat black holes with
a deficit solid angle and stealth solutions.

In addition to the traditional Einstein-Hilbert term, Langlois
\cite{langlois2021quadratic} found a surprisingly simple
reformulation of quadratic DHOST theories. These theories are based
on a Lagrangian, including a few geometrical terms related to the
three-dimensional constant hypersurfaces.
Boumaza and Langlois \cite{boumaza2022neutron} investigated the
neutron stars in the context of DHOST theories. In order to explain
the equation of state of the neutron star matter, they employed many
equations of state. In each instance, they numerically
estimated the neutron star profile using arbitrary values of energy
density and modified gravity parameters.

Frusciante \emph{et al.} \cite{frusciante2019tracker} examined
quadratic DHOST theories that meet degeneracy conditions. He chose
this theory to avoid Ostrogradsky instability, constraining GW speed
and determining the bound on the decay of GW to dark energy
perturbations. Furthermore, they generated the most generic
Lagrangians capable of reproducing tracking and scaling
characteristics individually. In the quadratic DHOST theories,
Takahashi and Motohashi \cite{takahashi2021black} investigated
perturbations regarding the stealth Schwarzschild-de-Sitter
solution. In addition to that, they investigated the sound speed in
the neighborhood of the black hole (BH). Subsequently, they
discovered that the gradient instability manifests itself in either
the radial or angular direction. Chen \emph {et al.}
\cite{chen2021testing} investigated the disformal Kerr BH frequency
in quadratic DHOST using the relativistic precession model.
Moreover, they found out that there is an almost negative disformal
parameter in the range of 1. They do so under some restrictions and
keeping in mind the GRO J1655-40 observational data.

On cosmic scales, the complexity factor is particularly significant
for self-gravitating systems. These systems exhibit a variety of
properties that have been extensively researched, including energy
density, pressure, stability, mass-radius ratio, and brightness. To
explore the complexity of the stellar structure the probability
distribution has been reformulated in the form of energy density
\cite{sanudo2009complexity,de2012entropy}. However, energy density
alone is insufficient to characterize complexity since the primary
factor pressure component of the energy-momentum tensor, which is
critical in the development of self-gravitating systems, is lacking
\cite{herrera2018new}.

Herrera \cite{herrera2009structure} conducted the study of
self-gravitating relativistic fluids with spherical symmetry in a
systematic manner. The approach relies on scalar functions that are
obtained from the orthogonal separation of the Riemann tensor.
According to this, the total number of scalars are five for
dissipative and anisotropic fluids, which can be reduced to two when
considering dissipationless dust or static anisotropic fluids.
Furthermore, only one scalar remains for static isotropic fluids.
Also, a new notion of complexity has been presented by him
\cite{herrera2018new} for relativistic fluid distributions with
static spherically symmetric properties. The underlying premise
assumes that a less complex system pertains to a homogeneous energy
density and isotropic pressure in the fluid. As an apparent
indicator of complexity, the scalar $Y_{TF}$ emerges as a viable
candidate for measurement purposes. Yousaf \emph{et al. }
\cite{yousaf2023quasi} examined the quasi-static evolution of
self-gravitating structures under strong curvature regimes,
considering higher-order curvature gravity and the Palatini
formalism of $f(R)$ gravity. Furthermore, the role of vorticity and
structure scalars in revealing fluid anisotropy effects and the
influence of Palatini-based factors on shearing motion is
highlighted. Eventually, they performed a comparison-based study to
explore the effects of distinct curvature factors on the propagation
of axial sources. Several researches has been done on this vast
topic of complexity by Herrera and different authors
\cite{lopez1995statistical,crutchfield2000comment,herrera2018definition,herrera2019complexity,herrera2019complexitya,
bhatti2021electromagnetic,jasim2021anisotropic,brassel2021inhomogeneous,gumede2021first,yousaf2022f,yousaf2023role}

Motivated by the work of Herrera \cite{herrera2009structure} and the
significance of scalar-tensor theories, we will construct the unique
notion of complexity within the context of DHOST theory. This
manuscript is organized in such a manner that Sec.~\textbf{II}
presents the comprehensive mathematical formalism of theory with
couplings between the scalar field and the Einstein tensor,
extracting the compact form of modified field equations in the
context of DHOST theory. In Sec.~\textbf{III}, we adapt a
well-suited spherically symmetric anisotropic matter content in
order to evaluate the modified field equations and non-conserved
equation. The major goal of our manuscript, i.e., to obtain the set
of structure scalars, is determined in Sec.~\textbf{IV}. Section~
\textbf{V} deals with the calculation of different masses. In Sec.~
\textbf{VI}, physical models under the constraint of minimal
complexity and the conformal flatness are discussed. Lastly, the
summary as well as the final outcomes of our investigations are
described in Sec.~\textbf{VII}. This manuscript will ultimately open
up the field of astrophysics to work in this modified gravity using
a variety of geometries. This study may assist in further
unravelling the enigma of the universe.

\section{BASIC FORMALISM OF DHOST THEORY}

\subsection{Motivation of DHOST Theories}

Scalar tensor theories have a history of being created in several
steps. Initially credited to Horndeski, higher-order theories were
built, leading to at most second-order Euler-Lagrange equations
(SOELE) for the metric and the scalar field. To get rid of the extra
scalar degree of freedom, it was mistakenly thought that at most
SOELE were required. As a result, Horndeski theories were thought to
be the most general theories free of Ostrogradsky instability. This
assumption was questioned by a modification of these theories that
are now frequently referred to as Beyond Horndeski (or GLPV) and
produced higher-order equations of motion. After, it was realized
that the degeneracy of their Lagrangian, rather than the order of
their equations of motion, is what distinguishes higher-order
theories propagating a single scalar degree of freedom, the Beyond
Horndeski theories were finally replaced by a larger class of
theories, the DHOST theories.

We investigated the structure of scalar-tensor theories of gravity
that are based on couplings between the derivative of scalar field
and matter degrees of freedom that are introduced by an effective
metric. Such interactions are categorized into disformal (vector),
conformal (scalar) and extended disformal (traceless tensor)
interactions according to their tensor structure. The absence of
Ostrogradski ghost degrees of freedom results from relations
restricted to the first derivative of the field, which guarantees
second-order equations of motion in the Einstein frame. These
derivative couplings play a crucial role in various cosmological
scenarios, such as inflation and dark energy models. They provide a
mechanism for the scalar field to interact with matter fields,
leading to interesting observational consequences. Furthermore, the
classification of these interactions based on their tensor structure
allows for a systematic study of their effects on the dynamics of
the universe.

These kind of Lagrangians have the significant the benefit of
containing numerous intriguing physical theories and enabling a
thorough examination of their properties. Such a broad approach has
been used to study the effective cosmological constant, inflationary
mechanism, compatibility with cosmological observations, and
cosmological dynamics. It is also important to remember that
Zumalac\'arregui and Garc\'ia-Bellido
\cite{zumalacarregui2014transforming} mentioned the opportunity to
develop these theories  by applying disformal transformations of the
metric to the Einstein-Hilbert action.

The Einstein frame representation of these theories introduces a
derivative coupling between the scalar field and the matter degrees
of freedom when the matter sector is taken into account. Important
phenomenological repercussions occurred from this. In high-density
environments, the scalar-mediated additional force can be concealed
by the disformal screening mechanism \cite{koivisto2012screening},
which is made possible by the derivative coupling. This effect could
be connected to the Vainshtein screening mechanism
\cite{vainshtein1972problem}, which conceals the scalar force within
a certain radius of point sources as a result of the non-linear
derivative self-interactions of the field brought on by the
degenerate terms $(Q(\phi,X)\Box
\phi,~\Sigma_{i=1}^{5}B_{i}(\phi,X)L_{i}^{(2)})$.
\\

\subsection{Formulations of DHOST theories}

The action integral for quadratic DHOST theories
\cite{langlois2016degenerate} is formulated as
\begin{align}\label{j1}
&S[\phi,g]=\int\sqrt{-g}\bigg\{R F(\phi,X)+B(\phi,X)+Q(\phi,X)\Box
\phi+\Sigma_{i=1}^{5}B_{i}(\phi,X)L_{i}^{(2)}+L_{m}\bigg\}d^{4}x,
\end{align}
where the functions $F,~B,~Q$ and $B_{i}$ depend on scalar field
$(\phi)$. In Eq. \eqref{j1}, $X=\nabla_{\mu}\phi\nabla^{\mu}\phi$,
$\phi_{\mu}$ is the gradient of scalar field, $R$ is the Ricci
scalar, $L_{i}$ are the multiple Lagrangians. Moreover, $L_{m}$ and
$g$ show matter Lagrangian and magnitude of metric tensor,
respectively. Equation \eqref{j1} can also be written as
\begin{align}\nonumber
&S[\phi,g]=\int\sqrt{-g}\bigg\{R F(\phi,X)+B(\phi,X)+Q(\phi,X)\Box
\phi+B_{1}(\phi,X)L_{1}^{(2)}+B_{2}(\phi,X)L_{2}^{(2)}
+B_{3}(\phi,X)L_{3}^{(2)}\\\label{j2}&+B_{4}(\phi,X)L_{4}^{(2)}
+B_{5}(\phi,X)L_{5}^{(2)}+L_{m}\bigg\}d^{4}x,
\end{align}
where
\begin{align}\nonumber
&L_{1}^{(2)}=\phi_{\pi\nu}\phi^{\pi\nu},\quad
L_{2}^{(2)}=(\Box\phi)^{2},\\\nonumber
&L_{3}^{(2)}=(\Box\phi)\phi^{\pi}\phi^{\nu}\phi_{\pi\nu},\quad
L_{4}^{(2)}=\phi^{\pi}\phi_{\pi\rho}\phi^{\rho\nu}\phi_{\nu},\\\nonumber
&L_{5}^{(2)}=(\phi_{\pi\nu}\phi^{\pi}\phi^{\nu})^{2}.
\end{align}
Now, we impose the shift symmetry which means that the DHOST action
is unchanged by the transformation $\phi\rightarrow\phi+c$ where $c$
is a constant. Thus, all the functions entering in the definition of
Eq. \eqref{j1} depend on $X$ only \cite{achour2020rotating}. Hence,
the final form of Eq. \eqref{j1} is
\begin{align}\nonumber
S[\phi,g]&=\int\sqrt{-g}\bigg\{R F(X)+B(X)+Q(X)\Box
\phi+B_{1}(X)\phi_{\pi\nu}\phi^{\pi\nu}
+B_{2}(X)(\Box\phi)^{2}+B_{3}(X)(\Box\phi)\phi^{\pi}\phi^{\nu}\phi_{\pi\nu}
\\\label{j3}&
+B_{4}(X)\phi^{\pi}\phi_{\pi\rho}\phi^{\rho\nu}\phi_{\nu}
+B_{5}(X)(\phi_{\pi\nu}\phi^{\pi}\phi^{\nu})^{2}+L_{m}\bigg\}d^{4}x.
\end{align}

The metric tensor components in the action integral play a crucial
role as gravitational potentials. By varying these components, we
can determine certain quantities that describe the system's nature
as a source of gravity and its ability to interact with a
gravitational field. In the context of General Relativity (GR), the
dynamical variable is represented by $g^{\mu\nu}$ (the metric
tensor). Consequently, to construct a valid Lagrangian density must
depend on the metric tensor. With the work of Brans and Dicke
\cite{brans1961mach}, a systematic study of scalar-tensor theories
as a modification of GR began. Subsequent developments mainly
focused on scalar-tensor Lagrangians with non-minimal couplings of
the scalar to curvature. Bearing in mind the approach adopted in
Brans-Dicke (BD) gravity, DHOST theories \cite{boumaza2022neutron}
and also the importance of the metric tensor, we vary the action
integral of DHOST theories concerning the metric tensor to achieve
its modified field equations. From another point of view, $X$ (the
kinetic term) also depends on the metric tensor, i.e.,
$X=g^{\alpha\beta}\nabla_{\alpha}\phi\nabla_{\beta}\phi$''.

Hence, by varying Eq. \eqref{j3} with respect to metric tensor, we
get
\begin{align}\nonumber
&\delta(S)=\int\bigg\{F\delta(\sqrt{-g}R)+\delta(\sqrt{-g})\bigg[
B+Q\Box
\phi+B_{1}\phi_{\alpha\beta}\phi^{\alpha\beta}+B_{2}(\Box\phi)^{2}
+B_{3}(\Box\phi)\phi^{\alpha}\phi^{\beta}\phi_{\alpha\beta}
+B_{4}\phi^{\lambda}\phi_{\lambda\rho}\phi^{\rho\gamma}\phi_{\gamma}\\\nonumber&
+B_{5}(\phi_{\alpha\beta}\phi^{\alpha}\phi^{\beta})^{2}\bigg]
+F_{X}\sqrt{-g}R+B_{X}\sqrt{-g}+Q_{X}\sqrt{-g}\Box\phi+Q\sqrt{-g}
(\partial_{\pi}\partial_{\nu}\phi)
\delta(g^{\pi\nu})+B_{1X}\phi_{\pi\nu}\phi^{\pi\nu}\sqrt{-g}\\\nonumber&
+B_{2X}\sqrt{-g}(\Box\phi)^{2}+2B_{2}\sqrt{-g}g^{\pi\nu}(\partial_{\pi}\partial_{\nu}\phi)^{2}
(\delta
g^{\pi\nu})+B_{3X}\sqrt{-g}(\Box\phi)\phi^{\alpha}\phi^{\beta}\phi_{\alpha\beta}
+B_{3}\sqrt{-g}\phi^{\alpha}\phi^{\beta}\phi_{\alpha\beta}\partial_{\pi}\partial_{\nu}\phi(\delta
g^{\pi\nu})\\\label{j4}&+B_{4X}\sqrt{-g}\phi^{\alpha}\phi_{\alpha\beta}\phi^{\beta\rho}\phi_{\rho}
+B_{5X}\sqrt{-g}(\phi_{\alpha\beta}\phi^{\alpha}\phi^{\beta})^{2}
+\delta(\sqrt{-g}L_{m})\bigg\}d^{4}x,
\end{align}
where $B_{X},~Q_{X}$ and $B_{iX}$ represents the derivative of $B,
~Q$ and $B_{i}'s$ with respect to kinetic term $X$, respectively.
Using the following formulas
\begin{align}\nonumber
\delta(\sqrt{-g})&=-\frac{1}{2}\sqrt{-g}g_{\pi\nu}\delta(g^{\pi\nu}),\\\nonumber
\delta(\sqrt{-g}R)&=\sqrt{-g}G_{\pi\nu}\delta(g^{\pi\nu}),\\\nonumber
T_{\pi\nu}&=g_{\pi\nu}L_{m}-2\frac{\partial L_{m}}{\partial
g^{\pi\nu}},
\end{align}
where $T_{\pi\nu}$ indicates the usual matter whose details will be
discussed later in this manuscript. After substituting these
formulae in Eq. \eqref{j4} and taking the terms
$\sqrt{-g}\delta(g^{\pi\nu})$ and $\sqrt{-g}$ common, we obtain
\begin{align}\nonumber
&\delta(S)=\int\bigg\{\sqrt{-g}\delta(g^{\pi\nu})\bigg[FG_{\pi\nu}-\frac{1}{2}g_{\pi\nu}
\bigg(B+Q\Box
\phi+B_{1}\phi_{\alpha\beta}\phi^{\alpha\beta}+B_{2}(\Box\phi)^{2}
+B_{3}(\Box\phi)\phi^{\alpha}\phi^{\beta}\phi_{\alpha\beta}\\\nonumber&
+B_{4}\phi^{\lambda}\phi_{\lambda\rho}\phi^{\rho\gamma}\phi_{\gamma}
+B_{5}(\phi_{\alpha\beta}\phi^{\alpha}\phi^{\beta})^{2}\bigg)
+Q\partial_{\pi}\partial_{\nu}\phi
+2B_{2}(g^{\alpha\beta}\nabla_{\alpha}\nabla_{\beta}\phi)
\partial_{\pi}\partial_{\nu}\phi
+B_{3}\phi^{\alpha}\phi^{\beta}\phi_{\alpha\beta}\partial_{\pi}\partial_{\nu}\phi\\\nonumber&-\kappa
T_{\pi\nu}\bigg] +\sqrt{-g}\bigg[F_{X}R+B_{X}+Q_{X}\Box\phi
+B_{1X}\phi_{\alpha\beta}\phi^{\alpha\beta}+B_{2X}(\Box\phi)^{2}
+B_{3X}(\Box\phi)\phi^{\alpha}\phi^{\beta}\phi_{\alpha\beta}
+B_{4X}\phi^{\alpha}\phi_{\alpha\beta}\phi^{\beta\rho}\phi_{\rho}
\\\nonumber
&+B_{5X}(\phi_{\alpha\beta}\phi^{\alpha}\phi^{\beta})^{2}\bigg]\bigg\}d^{4}x.
\end{align}
Next, multiplying and dividing the terms, having $\sqrt{-g}$ common,
with $-\delta(g^{\pi\nu})$.

After the multiplication and division of $-\delta(g^{\pi\nu})$, the
term in the numerator will be taken common but to further proceed
with our calculations and apply principle of least action, we have
to use the formula, $h^{\mu\nu}=-\delta g^{\mu\nu}$, in the
denominator. Eventually, we get
\begin{align}\nonumber
\delta(S)&=\int\bigg\{\sqrt{-g}\delta(g^{\pi\nu})\bigg[FG_{\pi\nu}-\frac{1}{2}g_{\pi\nu}
\bigg(B+Q\Box
\phi+B_{1}\phi_{\alpha\beta}\phi^{\alpha\beta}+B_{2}(\Box\phi)^{2}
+B_{3}(\Box\phi)\phi^{\alpha}\phi^{\beta}\phi_{\alpha\beta}\\\nonumber&
+B_{4}\phi^{\lambda}\phi_{\lambda\rho}\phi^{\rho\gamma}\phi_{\gamma}
+B_{5}(\phi_{\alpha\beta}\phi^{\alpha}\phi^{\beta})^{2}\bigg)+Q\partial_{\pi}\partial_{\nu}\phi
+2B_{2}(g^{\alpha\beta}\nabla_{\alpha}\nabla_{\beta}\phi)\partial_{\pi}\partial_{\nu}\phi
+B_{3}\phi^{\alpha}\phi^{\beta}\phi_{\alpha\beta}\partial_{\pi}\partial_{\nu}\phi\\\nonumber&-\kappa
T_{\pi\nu}\bigg]-\frac{\sqrt{-g}\delta(g^{\pi\nu})}{h^{\pi\nu}}\bigg[F_{X}R+B_{X}+Q_{X}\Box\phi
+B_{1X}\phi_{\alpha\beta}\phi^{\alpha\beta}+B_{2X}(\Box\phi)^{2}
+B_{3X}(\Box\phi)\phi^{\alpha}\\\label{j5}&\phi_{\alpha\beta}\phi^{\beta}
+B_{4X}\phi^{\alpha}\phi_{\alpha\beta}\phi^{\beta\rho}\phi_{\rho}
+B_{5X}(\phi_{\alpha\beta}\phi^{\alpha}\phi^{\beta})^{2}\bigg]\bigg\}d^{4}x.
\end{align}
According to the variational principle, we have $\delta(S)=0$. As
the integral cannot be equal to zero so the integrand will be equal
to zero and finally we are left with
\begin{align}\nonumber
&FG_{\pi\nu}-\frac{1}{2}g_{\pi\nu}\bigg(B+Q\Box
\phi+B_{1}\phi_{\alpha\beta}\phi^{\alpha\beta}+B_{2}(\Box\phi)^{2}
+B_{3}(\Box\phi)\phi^{\alpha}\phi^{\beta}\phi_{\alpha\beta}
+B_{4}\phi^{\lambda}\phi_{\lambda\rho}\phi^{\rho\gamma}\phi_{\gamma}
+B_{5}(\phi_{\alpha\beta}\phi^{\alpha}\phi^{\beta})^{2}\bigg)\\\nonumber&
+Q\partial_{\pi}\partial_{\nu}\phi
+2B_{2}(g^{\alpha\beta}\nabla_{\alpha}\nabla_{\beta}\phi)\partial_{\pi}\partial_{\nu}\phi
+B_{3}\phi^{\alpha}\phi^{\beta}\phi_{\alpha\beta}\partial_{\pi}\partial_{\nu}\phi-\kappa
T_{\pi\nu}-\frac{1}{h^{\pi\nu}}\bigg[F_{X}R+B_{X}+Q_{X}\Box\phi
+B_{1X}\phi_{\alpha\beta}\phi^{\alpha\beta}\\\label{j6}&
+B_{3X}(\Box\phi)\phi^{\alpha}\phi^{\beta}\phi_{\alpha\beta}+B_{2X}(\Box\phi)^{2}
+B_{4X}\phi^{\alpha}\phi_{\alpha\beta}\phi^{\beta\rho}\phi_{\rho}
+B_{5X}(\phi_{\alpha\beta}\phi^{\alpha}\phi^{\beta})^{2}\bigg]=0.
\end{align}
Rearranging Eq. \eqref{j6}, we accomplish
\begin{align}\nonumber
&G_{\pi\nu}=\frac{1}{F}\bigg\{\kappa
T_{\pi\nu}+\frac{1}{2}g_{\pi\nu}\bigg(B+Q\Box
\phi+B_{1}\phi_{\alpha\beta}\phi^{\alpha\beta}+B_{2}(\Box\phi)^{2}
+B_{3}(\Box\phi)\phi^{\rho}\phi^{\lambda}\phi_{\rho\lambda}
+B_{4}\phi^{\lambda}\phi_{\lambda\rho}\phi^{\rho\gamma}\phi_{\gamma}\\\nonumber&
+B_{5}(\phi_{\rho\lambda}\phi^{\rho}\phi^{\lambda})^{2}\bigg)-Q\partial_{\pi}\partial_{\nu}\phi
-2B_{2}(g^{\alpha\beta}\nabla_{\alpha}\nabla_{\beta}\phi)\partial_{\pi}\partial_{\nu}\phi
-B_{3}\phi^{\alpha}\phi^{\beta}\phi_{\alpha\beta}\partial_{\pi}\partial_{\nu}\phi
+\frac{1}{h^{\pi\nu}}\bigg[F_{X}R+B_{X}+Q_{X}\Box\phi\\\label{j6*}&
+B_{1X}\phi_{\alpha\beta}\phi^{\alpha\beta}+B_{2X}(\Box\phi)^{2}
+B_{3X}(\Box\phi)\phi^{\alpha}\phi^{\beta}\phi_{\alpha\beta}
+B_{4X}\phi^{\alpha}\phi_{\alpha\beta}\phi^{\beta\rho}\phi_{\rho}
+B_{5X}(\phi_{\alpha\beta}\phi^{\alpha}\phi^{\beta})^{2}\bigg]\bigg\}.
\end{align}
In DHOST theories, the constant kinetic term refers to a specific
aspect of the action. It is represented by a consistent coefficient
that multiplies the kinetic term of the scalar field. The role of
this kinetic term is to capture the behaviour associated with second
derivatives of the scalar field within the action formulation. The
inclusion of a constant kinetic term in DHOST theories carries
significant theoretical and practical implications for the behaviour
of scalar fields and gravitational interactions. This constant
kinetic term plays a crucial role in determining the dynamics of
both the gravity and scalar field. The presence of this term
contributes to shaping important physical phenomena and can impact
various aspects related to the scalar field's evolution and its
interaction with gravitational forces.

The presence of constant kinetic terms has considerable implications
for the dynamics of cosmology. It introduces changes in the rate at
which the universe expands, as well as affecting the growth patterns
of structures and other observable phenomena in cosmology.
Additionally, this constant kinetic term can exert an influence on
how scalar fields interact under different conditions or scenarios.
Notably, certain situations may trigger screening mechanisms
\cite{vainshtein1972problem,koivisto2012screening} that suppress any
impact from these scalar fields on small scales or highly dense
regions, making their behaviour consistent with various
observational constraints. The assumption that $X$ is a constant,
specifically $X=X_{0}$, greatly simplifies the modified Einstein
equations. This also has led many researchers to explore stealth
solutions
\cite{takahashi2019linear,achour2020rotating,babichev2020regular,takahashi2021black}.

Hence, modified Einstein equations are enormously simplified if one
considers the solution as $X=X_{0}$, where $X_{0}$ is a constant. By
implementing the aforementioned assumption on Eq. \eqref{j6*} all
the derivatives of coupling functions with respect to the kinetic
term will vanish. Then Eq. \eqref{j6*} is accomplished as
\begin{align}\nonumber
&G_{\pi\nu}=\frac{1}{F}\bigg\{\kappa
T_{\pi\nu}+\frac{1}{2}g_{\pi\nu}\bigg(B+Q\Box
\phi+B_{1}\phi_{\alpha\beta}\phi^{\alpha\beta}+B_{2}(\Box \phi)^{2}
+B_{3}\Box \phi\phi^{\rho}\phi^{\lambda}\phi_{\rho\lambda}
+B_{4}\phi^{\lambda}\phi_{\lambda\rho}\phi^{\rho\gamma}\phi_{\gamma}
\\\label{j7}&+B_{5}(\phi_{\rho\lambda}\phi^{\rho}\phi^{\lambda})^{2}\bigg)-Q\partial_{\pi}\partial_{\nu}\phi
-2B_{2}\Box \phi\partial_{\pi}\partial_{\nu}\phi
-B_{3}\phi^{\alpha}\phi^{\beta}\phi_{\alpha\beta}\partial_{\pi}\partial_{\nu}\phi\bigg\}.
\end{align}
Equation \eqref{j7}, in the compact form, can be penned as
\begin{align}\label{j7*}
&G_{\pi\nu}=T_{\pi\nu}^{(M)},
\end{align}
where the value of $T_{\pi\nu}^{(M)}$ is
\begin{align}\nonumber
T_{\pi\nu}^{(M)}&=\frac{1}{F}\bigg\{\kappa
T_{\pi\nu}+\frac{1}{2}g_{\pi\nu}\bigg(B+Q\Box
\phi+B_{1}\phi_{\alpha\beta}\phi^{\alpha\beta}+B_{2}(\Box\phi)^{2}
+B_{3}(\Box\phi)\phi^{\rho}\phi^{\lambda}\phi_{\rho\lambda}
+B_{4}\phi^{\lambda}\phi_{\lambda\rho}\phi^{\rho\gamma}\phi_{\gamma}
\\\nonumber&+B_{5}(\phi_{\rho\lambda}\phi^{\rho}\phi^{\lambda})^{2}\bigg)
-Q\partial_{\pi}\partial_{\nu}\phi
-2B_{2}(g^{\alpha\beta}\nabla_{\alpha}\nabla_{\beta}\phi)\partial_{\pi}\partial_{\nu}\phi
-B_{3}\phi^{\alpha}\phi^{\beta}\phi_{\alpha\beta}\partial_{\pi}\partial_{\nu}\phi\bigg\}.
\end{align}
Now, we discuss about the usual matter. We assume it as follows
\begin{align}\label{j8}
T_{\pi\nu}=&(\rho+P)V_\pi V_\nu-Pg_{\pi\nu}+\Pi_{\pi\nu},
\end{align}
where $\rho,~\Pi_{\pi\nu}$ and $P$ express energy density,
anisotropic tensor and pressure, respectively. Further
\begin{align}\nonumber
&P=\frac{2P_{\bot}+P_{r}}{3},\quad \Pi_{\pi\nu}=\Pi\bigg(K_\pi
K_\nu-\frac{h_{\pi\nu}}{3}\bigg),\\\nonumber
&h_{\pi\nu}=g_{\pi\nu}+V_{\nu}V_{\pi},\quad \Pi=P_{r}-P_{\bot},
\end{align}
where pressure is exerted in two particular directions that are
termed as radial $P_{r}$ and tangential $P_{\perp}$ pressures. Here,
$K_{\pi}$ represents the orthonormal basis \cite{van1997general} and
$h_{\pi\nu}$ is the projection tensor that projects the timelike
vector into the (spacelike) subspace orthogonal to the four-velocity
$V$. Sets of vector fields that are locally defined, orthonormal to
one another, and linearly independent are known as orthonormal
bases. The definition of a locally inertial frame and the
description of the tensor components as perceived by an observer at
rest in that frame are made easier if orthonormal bases are present.
These bases, which offer significantly more benefits than the
coordinate basis, are frequently employed in the literature
\cite{herrera2021hyperbolically}.

\section{Gravitational Equations}
We take into account an anisotropic matter content with spherically
symmetric metric as
\begin{equation}\label{j8a}
ds^2=-e^{\lambda(r)}dr^2-r^2(d\theta^2+\sin^2 \theta
d\phi^2)+e^{\nu(r)}dt^2,
\end{equation}
where $\lambda$ and $\nu$ are functions of $r$. The field equations
in DHOST theory are evaluated using Eqs. \eqref{j7} and \eqref{j8a}
as
\begin{align}\label{j9}
&\frac{\lambda^{\prime}r+e^\lambda-1}{e^\lambda r^2}=\frac{8 \pi
\rho}{F}+\frac{1}{2 F}\bigg[B+Q \gamma_1+B_1
\gamma_2+B_2\gamma_3+B_3 \gamma_4+B_4 \gamma_5+B_5
\gamma_6\bigg],\\\label{j10}&\frac{1-e^{-\lambda}-e^{-\lambda}
\nu^{\prime}r}{r^2}=\frac{-8 \pi P_r}{F}+\frac{1}{2F} \bigg[B+Q
\gamma_{7}+B_{1}
\gamma_{2}+B_{2}\gamma_{8}+B_{3}\gamma_{9}+B_{4}\gamma_{5}+B_{5}
\gamma_{6}\bigg],\\\nonumber & \frac{1}{4
e^\lambda}\bigg[\lambda^{\prime} \nu^{\prime}-\nu^{\prime 2}-2
\nu^{\prime \prime}+\frac{2 \lambda^{\prime}}{r} -\frac{2
\nu^{\prime}}{r}\bigg]=-\frac{8\pi
P_\perp}{F}+\frac{1}{2F}\bigg[B+Q\gamma_1+B_{1} \gamma_2+B_{2}
\gamma_{3}+B_{3} \gamma_{10}\\\label{j11}&+B_{4} \gamma_{5} +B_{5}
\gamma_{6}\bigg],
\end{align}
where the values of $\gamma_{i}'s$ $(i=1-10)$ are given in Appendix.
Moreover, if one works within the Einstein frame, where the dilaton
emerges as part of the matter sector, non-conservation can also be
attained, similar to Brans-Dicke theory, which is popular in some
scalar-tensor theories \cite{brans1961mach,velten2021conserve}. One
can define the dilaton as a scalar field that arises in certain
gravitational theories, most notably string theory and some extended
modified theories which has considerable consequences for the
dynamics of spacetime. Subsequently, we have obtained the
non-conserved equation using the expression
$T^{(M)}_{\pi\nu;\nu}\neq0$ (we assume it to be $Z$)
\begin{align}\nonumber
&-\frac{\nu^{\prime}}{2F}\bigg[(\rho-P_{r})+\frac{1}{8\pi}\bigg(B+2
Q(\gamma_{1}+\gamma_{7})+B_{1}\gamma_{2}+B_{2}(\gamma_{3}+\gamma_{8})+
B_{3}(\gamma_{4}+\gamma_{9})+B_{4}\gamma_{5}+B_{5}\gamma_{6}
\bigg)\bigg]\\\nonumber&+\bigg(\frac{P_{r}}{F}\bigg)'-\bigg[\frac{1}{16\pi
F}\bigg(B+Q\gamma_{7}+B_{1}\gamma_{2}
+B_{2}\gamma_{8}+B_{3}\gamma_{9}+B_{4} \gamma_{5}+B_{5} \gamma_{6}
\bigg)\bigg]'+\frac{2 \Pi}{rF}-\frac{1}{16\pi
rF}\bigg[Q(\gamma_{7}\\\label{j12}&-\gamma_{1})+B_2(\gamma_{8}-\gamma_{3})+B_{3}(\gamma_{9}
-\gamma_{10})\bigg]=-Ze^{\lambda}.
\end{align}
The Tolman-Oppenheimer-Volkoff (TOV) equation is a result of the
conservation of energy in stellar objects. This equation relates the
composition of stars to their metric. Moreover, it has gained
considerable fame in astrophysics for its significance in
understanding stellar dynamics. In our scenario, due to the presence
of the ADF in DHOST theory, it becomes non-conserved.

DHOST theories are a broad class of modified gravity theories which
can exhibit a variety of conservation properties, depending on their
formulation and the context in which they are considered. The
conservation of energy-momentum in gravitational interactions can be
preserved in some DHOST theory formulations, which is analogous to
GR. These hypotheses are regarded as conserved. DHOST theories,
however, may result in violations of energy-momentum conservation in
other formulations, leading to non-conserved theories. In these
circumstances, the addition of higher-order derivative terms may
have an impact on the scalar field's dynamics and its interactions
with gravity, causing deviations from the conservation laws of
standard GR.

\section{Structure scalars}

Herrera \emph{et al. } \cite{herrera2009structure} calculated the
structure scalars and created a comprehensive set of equations.
These equations describe the structure of anisotropic
self-gravitating and spherically symmetric matter content based on
five scalar quantities $(Y_{T},~Y_{TF},~X_{T},~X_{TF},~\mathbb{Z})$.
These scalar quantities constructed from the orthogonal splitting of
Curvature tensor in GR. In this section, we will obtain the
structure scalars using the aforementioned approach. For this
purpose, we begin with the equation relating the Weyl tensor, Ricci
scalar, and intrinsic curvature is given as
\begin{align}\nonumber
&R^\pi_{\gamma\nu\delta}=C^\pi_{\gamma\nu\delta}+\frac{1}{2}R^\pi_\nu
g_{\gamma\delta}+\frac{1}{2}R_{\gamma\nu}\delta^\pi_\delta+\frac{1}{2}R_{\gamma\delta}\delta^\pi_\nu
\\\label{j13}&-\frac{1}{2}R^\pi_\delta
g_{\gamma\nu}-\frac{1}{6}R\bigg(\delta^\pi_\nu
g_{\gamma\delta}-g_{\gamma\nu}\delta^\pi_\delta\bigg),
\end{align}
where $C^\pi_{\gamma\nu\delta}$ is defined as the Weyl tensor and is
composed of electric and magnetic parts. Due to the spherical
system, the magnetic part vanishes and the electric part can be
written as
\begin{equation}\label{j14}
E_{\pi\nu}=\varepsilon\left(K_{\pi} K_{\nu}+\frac{1}{3}
h_{\pi\nu}\right),
\end{equation}
where conformal scalar, i.e., $\varepsilon$ is calculated for our
metric as
\begin{align}\nonumber
\varepsilon&=-\frac{{\nu}''e^{-\lambda}}{4}+\frac{\lambda'\nu'e^{-\lambda}}{8}-\frac{\nu'^2
e^{-\lambda}}{8}+\frac{\nu'e^{-\lambda}}{4r}\\\label{j15}&-
\frac{\lambda'e^{-\lambda}}{4r}-\frac{e^{-\lambda}}{2r^2}-\frac{1}{2r^2}.
\end{align}
Let us now present the following tensors \cite{herrera2009structure}
\begin{eqnarray}\label{j16}
&&Y_{\pi\nu}=^*R_{\pi\eta\nu\xi}V^\eta V^\xi,\\\label{j17}
&&\mathbb{Z}_{\pi\nu}=^*R_{\pi\eta\nu\xi}V^\eta
V^\xi=\frac{1}{2}\eta_{\pi\eta\epsilon\rho}
{^*R^{\epsilon\rho}_{\nu\xi}}V^\eta V^\xi,\\\label{j18}
&&X_{\pi\nu}=^*R^*_{\pi\eta\nu\xi}V^\eta
V^\xi=\frac{1}{2}\eta_{\pi\eta}^{\epsilon\rho}R^*_{\epsilon\rho\nu\xi}V^\eta
V^\xi.
\end{eqnarray}
With the help of modified field equations, one can write Eq.
\eqref{j13} as
\begin{align}\label{j18*}
R^{\nu\delta}_{\pi\gamma}&=C^{\nu\delta}_{\pi\gamma}+16\pi
\left.T^{(M)}\right.^{[\nu}_{[\pi}\delta^{\delta]}_{\gamma]}+8\pi
\left.T^{(M)}\right.\bigg(\frac{1}{3}\delta^\nu_{[\pi}\delta^\delta_{\gamma]}-\delta^{[\nu}_{[\pi}\delta^{\delta]}_{\gamma]}
\bigg),
\end{align}
where the value of
$\left.T^{(M)}\right.^{[\nu}_{[\pi}\delta^{\delta]}_{\gamma]}$ is
defined in the Appendix. The splitting of the Riemann tensor $R_{\nu
\delta}^{\pi \gamma}$ gives
\begin{equation}\nonumber
R^{\pi\gamma}_{\nu\delta}=R^{\pi\gamma}_{(I)\nu\delta}
+R^{\pi\gamma}_{(II)\nu\delta}+R^{\pi\gamma}_{(III)\nu\delta},
\end{equation}
and utilizing Eq. \eqref{j18*}, we obtain
\begin{align}\nonumber
R_{(I)\nu \delta}^{\pi \gamma}&=\frac{16\pi}{F}\bigg\{\rho V^{[\pi}
V_{[\nu} \delta_{\delta]}^{\gamma]}-P h_{[\nu}^{[\pi}
\delta_{\delta]}^{\gamma]}\bigg\}+\frac{8}{3
F}\bigg[\pi(\rho+3P)+2I_{1}-Q\Box \phi+2 B_{2}(\Box
\phi)^2-B_{3}\phi^\alpha \phi^\beta \phi_{\alpha \beta} \Box
\phi\bigg]\\\label{j19}&\times\bigg[\delta_{\delta}^{\pi}
\delta_{\nu}^{\gamma}-\delta_{\nu}^{\pi}
\delta_{\delta}^{\gamma}\bigg], \\\nonumber R_{(II)\nu \delta}^{\pi
\gamma}&=\frac{16\pi }{F}\Pi_{[\nu}^{[\pi}
\delta_{\delta]}^{\gamma]}+\frac{1}{2}\bigg\{\frac{I_{1}}{2F}\bigg(\delta_{\gamma}^{\pi}
\delta_{\delta}^{\nu}-\delta_{\nu}^{\gamma}
\delta_{\delta}^{\pi}+\delta_{\delta}^{\gamma}
\delta_{\nu}^{\pi}-\delta_{\delta}^{\pi}
\delta_{\nu}^{\gamma}\bigg)-Q\bigg(\delta_{\delta}^{\gamma}(\partial^{\pi}\partial_{\nu}
\phi)-\delta_{\nu}^{\gamma}(\partial^{\pi}\partial_{\delta}\phi)
-\delta_{\delta}^{\pi}(\partial^{\gamma}\partial_{\nu}
\phi)\\\nonumber&+\delta_{\nu}^{\pi}(\partial^{\gamma}
\partial_{\delta}\phi)\bigg)-2B_{2}\Box \phi
\bigg(\delta_{\delta}^{\gamma}(\partial^{\pi}\partial_{\nu}
\phi)-\delta_{\nu}^{\gamma}(\partial^{\pi}\partial_{\delta}\phi)
-\delta_{\delta}^{\pi}(\partial^{\gamma}\partial_{\nu}
\phi)+\delta_{\nu}^{\pi}(\partial^{\gamma}
\partial_{\delta}\phi)\bigg)-B_3 \phi^\alpha \phi^\beta
\phi_{\alpha
\beta}\\\label{j20}&\times\bigg(\delta_{\delta}^{\gamma}(\partial^{\pi}\partial_{\nu}
\phi)-\delta_{\nu}^{\gamma}(\partial^{\pi}\partial_{\delta}\phi)
-\delta_{\delta}^{\pi}(\partial^{\gamma}\partial_{\nu}
\phi)+\delta_{\nu}^{\pi}(\partial^{\gamma}
\partial_{\delta}\phi)\bigg)\bigg\},
\\\label{j21} R_{(III)\nu \delta}^{\pi \gamma}&=4 V^{[\pi} V_{[\nu}
E_{\delta]}^{\gamma]} -\epsilon_\eta^{\pi \gamma}
\epsilon_{\xi\nu\delta} E^{\eta\xi},
\end{align}
where $I_{1}$ is defined in Appendix. After this, one can easily
evaluate the tensors $Y_{\pi\nu},~X_{\pi\nu}$ and $Z_{\pi\nu}$ in
terms of state variables using Eqs. \eqref{j16}-\eqref{j21} as
\begin{align}\nonumber
Y_{\pi\nu}&=E_{\pi \nu}+\frac{4\pi h_{\pi
\nu}}{F}(\rho-P)+\frac{h_{\pi \nu}}{2F}I_{1} +\bigg[\partial_{\pi}
\partial_{\nu} \phi-(\partial_{\pi}
\partial_{\alpha} \phi) V_{\nu} V^{\alpha}-V_{\pi}
V_{\alpha}(\partial^{\alpha} \partial_{\nu} \phi)+g_{\pi\nu}
V_\alpha V^{\beta}(\partial^{\alpha}
\partial_{\beta}
\phi)\bigg]\\\label{j22}&\times\bigg[-Q-2B_{2}\Box
\phi-B_{3}\phi^{\alpha} \phi^{\beta}
\phi_{\alpha\beta}\bigg]-\frac{8h_{\pi\nu}}{3F}\bigg[\pi(\rho+3P)+2I_{1}-Q
\Box \phi -2B_{2}(\Box \phi)^{2}-B_{2}\phi^{\alpha}
\phi^{\beta}\phi_{\alpha\beta}\Box \phi\bigg],\\\nonumber
X_{\pi\nu}&=-E_{\pi\nu}+\frac{8 \pi}{3 F}\rho
h_{\pi\nu}+\frac{4\pi\Pi_{\pi\nu}}{F}-\frac{I_{1}}{2F}
h_{\pi\nu}+\frac{1}{4}\bigg[(\partial^{\alpha}\partial_{\beta}\phi)
\epsilon_{\pi}^{\beta\delta}
\epsilon_{\alpha\delta\nu}-(\partial^{\alpha} \partial_{\delta}\phi)
\epsilon_{\pi}^{\beta\delta}
\epsilon_{\alpha\beta\nu}-(\partial^{\gamma} \partial_{\beta}\phi)
\epsilon_{\pi}^{\beta \delta} \epsilon_{\delta\gamma \nu}\\\nonumber
&+(\partial^{\gamma}
\partial_{\delta}\phi) \epsilon_{\pi}^{\beta\delta} \epsilon_{\beta
\gamma \nu}\bigg]\bigg[-Q-2B_{2}\Box
\phi-B_{3}\phi^{\alpha}\phi^{\beta}\phi_{\alpha\beta}\bigg]
+\frac{8}{3 F}\bigg[2I_{1}-Q\Box \phi-2B_{2}(\Box
\phi)^{2}-B_{2}\phi^{\alpha}\phi^{\beta} \phi_{\alpha\beta}\Box
\phi\bigg]\\\label{j23}&\times h_{\pi\nu},
\\\label{j24} \mathbb{Z}_{\pi
\nu}&=\frac{-1}{2 F}\bigg[Q_1+2 B_2\Box \phi+B_3 \phi^\alpha
\phi^\beta \phi_{\alpha
\beta}\bigg](\partial^{\gamma}\partial_{\delta}\phi)\epsilon_{\nu\gamma\pi}V^{\delta}.
\end{align}
In order to write the tensors $X_{\pi\nu}$ and $Y_{\pi\nu}$ in the
form of trace and trace free parts, we use
\cite{herrera2018definition}
\begin{align}\nonumber
X_{\pi\nu}&=\frac{h_{\pi\nu}}{3}X_T +\left(K_\pi K_\nu
+\frac{h_{\pi\nu}}{3}\right)X_{TF},\\\nonumber
Y_{\pi\nu}&=\frac{h_{\pi\nu}}{3}Y_T +\left(K_\pi K_\nu
+\frac{h_{\pi\nu}}{3}\right)Y_{TF}.
\end{align}
The aforementioned formulae help us to accomplish the trace and
trace free parts for tensor $Y_{\pi\nu}$ as
\begin{align}\nonumber
Y_{T}&=\frac{4\pi}{F}(\rho+3P)-\frac{3I_{1}}{2F} \bigg[\Box
\phi-(\partial_{\pi}\partial_{\delta} \phi) V^{\pi}
V^{\delta}-V^{\nu} V_{\gamma}(\partial_{\nu}^{\gamma}\phi)
+4V_{\gamma} V^{\delta}(\partial^{\gamma}
\partial_{\delta}\phi)\bigg]
\bigg[-Q-2B_{2}\Box
\phi\\\label{j25}&-B_{3}\phi^{\alpha}\phi^{\beta} \phi_{\alpha
\beta}\bigg]-\frac{8}{F}\bigg[2I_{1}-Q\Box \phi -2B_{2}(\Box
\phi)^{2}-B_{3}\phi^{\alpha} \phi^{\beta} \phi_{\alpha\beta}\Box
\phi\bigg],\\\label{j26} Y_{TF}&=\varepsilon+\frac{4\pi
\Pi}{F}+\frac{I_{2}}{K_\pi K_\nu +\frac{h_{\pi\nu}}{3}},
\end{align}
where $I_{2}$ is defined in Appendix.

A novel notion of intricacy has been presented in the context of
static spherically symmetric relativistic fluid distributions. The
assumption underlying this concept says that a less intricate system
is characterized by a homogeneous energy density and isotropic
pressure. Therefore, to gauge the level of intricacy, the structure
scalar $Y_{T F}$ emerges as an obvious choice \cite{herrera2018new}.
Following this $Y_{TF}$ is termed as the complexity factor (CF) in
our scenario as it gives the maximum information about the galactic
structures. In addition to that, the trace and trace-free portion of
the tensor $X_{\pi\nu}$ are
\begin{align}\nonumber
& X_{T}=\frac{8\pi\rho}{F}-\frac{3I_{1}}{2F}+\frac{1}{2}
\bigg[Q+2B_{2}\Box \phi+B_{3}\phi^{\alpha}\phi^{\beta}\phi_{\alpha
\beta}\bigg]\bigg[h_\alpha^{\beta}\partial ^{\alpha}\partial_{\beta}
\phi+h_{\alpha}^{\delta}(\partial
^{\alpha}\partial_{\delta}\phi)+h_{\gamma}^{\beta}(\partial^{\gamma}
\partial_{\beta} \phi)\\\label{j27}&+h_{\gamma}^{\delta}
(\partial^{\gamma}\partial_{\delta}\phi)\bigg]+\frac{8}{F}\bigg[2I_{1}-Q\Box
\phi-2 B_{2}(\Box
\phi)^2-B_{3}\phi^{\alpha}\phi^{\beta}\phi_{\alpha\beta}\Box
\phi\bigg],\\\label{j28} X_{TF}&=-\varepsilon+\frac{4\pi
\Pi}{F}+\frac{I_{3}}{K_\pi K_\nu +\frac{h_{\pi\nu}}{3}},
\end{align}
where $I_{3}$ is defined in Appendix.

Equations \eqref{j26} and \eqref{j28} produce
\begin{align}\nonumber
X_{TF}+Y_{TF}&=\frac{8\pi\Pi}{F}+\frac{I_{2}+I_{3}}{K_{\pi}K_{\nu}+\frac{h_{\pi\nu}}{3}}.
\end{align}
The development and growth of galactic structures are significantly
influenced by realistic parameters including pressure anisotropy and
energy density inhomogeneity. The structure scalars are significant
in describing the physical characteristics of stellar configurations
in modern relativistic astrophysics. The characteristics of the
suggested structures must be understood to calculate the CF of the
celestial bodies. These elements characterize the degree of
structure complexity and serve as a standard for comparing different
stellar configurations. The need to ascertain the complexity of
celestial objects is not new, but there have been several attempts
to accomplish this
\cite{herrera2019complexity,andrade2022anisotropic,yousaf2022structure}.

\section{Misner-Sharp and Tolman mass}

The concept of defining energy within a closed surface can be quite
challenging. However, in the specific case of spherical symmetry,
there exists a method to quantify the mass energy inside a sphere
known as the Misner-Sharp mass function. This allows for a
quantification of mass energy within spherical symmetry and provides
valuable insights into gravitational systems. This mass is
particularly useful for analyzing gravitational collapse in
spherically symmetric spacetimes and can be utilized to define
physical mass energy. Now, we utilize the Misner-Sharp
\cite{misner1964relativistic} mass to describe the matter content of
geometry and gravity. It is obtained by applying mass transition
from the matter to the gravitational area through the collapsing
process. It may be claimed that this mass is the "quasi-local"
version of mass because it is defined at the borders of a particular
region of spacetime. The mathematical expression for this mass is
\begin{equation}\label{j29}
m(r)=\frac{r}{2}(1-e^{-\lambda}).
\end{equation}
With the support of Eqs. \eqref{j9} and \eqref{j29}, one can
accomplish
\begin{align}\label{j30}
&m=4\pi\int\frac{r^2 \rho}{F}dr+\int \frac{r^2}{4F} \bigg[B+Q
\gamma_{1}+B_{1}\gamma_{2}+B_{2}\gamma_{3}+B_{3}\gamma_{4}
+B_{4}\gamma_{5}+B_{5}\gamma_{6}\bigg]dr.
\end{align}
The combination of Eqs. \eqref{j10} and \eqref{j29} produce
\begin{align}\label{j31}
&\nu^{\prime}=\frac{8 \pi r^3 P_r}{Fr(r-2 m)}+\frac{1}{16 \pi Fr(r-2
m)} \bigg(B+Q \gamma_7+B_1 \gamma_2+B_2 \gamma_8+B_3 \gamma_9+B_4
\gamma_5+B_5 \gamma_6\bigg)+2m.
\end{align}
Using the obtained value of $\nu'$ in the non-conserved equation, we
accomplish
\begin{align}\nonumber
&\bigg(\frac{P_r}{F}\bigg)'-\bigg[\frac{8 \pi r^3 P_r}{2F^{2}r(r-2
m)}+\frac{1}{32\pi F^{2}r(r-2 m)}\bigg(B+Q \gamma_7+B_1 \gamma_2+B_2
\gamma_8+B_3 \gamma_9+B_4 \gamma_5+B_5 \gamma_6\bigg)+2
m\bigg]\\\nonumber&\times\bigg[(\rho-P)+\frac{1}{8\pi}\bigg(B+2
Q(\gamma_{1}+\gamma_{7})+B_{1}\gamma_{2}+B_{2}(\gamma_{3}+\gamma_{8})+2
B_{3}(\gamma_{4}+\gamma_{9})+B_{4}\gamma_{5}+B_{5}\gamma_{6}\bigg)\bigg]-
\bigg[\frac{1}{16\pi
F}\\\nonumber&\times\bigg(B+Q\gamma_{7}+B_{2}\gamma_{2}
+B_{2}\gamma_{3}+B_{3}\gamma_{8}+B_{4} \gamma_{5}+B_{5} \gamma_{6}
\bigg)\bigg]'+\frac{2 \Pi}{rF}-\frac{1}{16\pi
rF}\bigg[Q(\gamma_{7}-\gamma_{1})+B_2(\gamma_{8}-\gamma_{3})\\\label{j32}&+B_{3}(\gamma_{9}
-\gamma_{10})\bigg]=-Ze^{\lambda}.
\end{align}
With the involvement of the Eqs. \eqref{j9}-\eqref{j11}, \eqref{j15}
and \eqref{j29}, one can achieve
\begin{align}\nonumber
&m=\frac{\varepsilon r^3}{3}+\frac{4 \pi r^3}{3 F}
(\rho+P_{\perp}-P_r)+\frac{r^3}{6
F}\bigg[B+Q(2\gamma_1-\gamma_7)+B_1 \gamma_2+B_2\left(2
\gamma_3-\gamma_{8}\right)
\\\label{j33}&+B_3\left(\gamma_4+\gamma_{10}-\gamma_{9}\right)+B_4 \gamma_5+B_5 \gamma_6\bigg].
\end{align}
The Weyl scalar can be determined in terms of density and pressure
using Eqs. \eqref{j30} and \eqref{j33} as
\begin{align}\nonumber
&\varepsilon=\frac{3}{r^3}\bigg[4 \pi \int \frac{r^2 \rho}{F} d r+
\int \frac{r^2}{4 F}[B+Q \gamma_1+B_1 \gamma_2+B_2 \gamma_3+B_3
\gamma_4+B_4 \gamma_5+B_5 \gamma_6] dr\bigg]-\frac{4 \pi
(\rho+P_{\perp}-P_r)}{F}\\\label{j34}&-\frac{1}{2
F}\bigg[B+Q(2\gamma_1-\gamma_7)+B_1 \gamma_2+B_2(2
\gamma_3-\gamma_{8})+B_4\gamma_5+B_3(\gamma_4+\gamma_{10}-\gamma_{9})+B_5
\gamma_6\bigg].
\end{align}
Next, we evaluate the Tolman mass \cite{tolman1930use}. It can be
formulated for the spherically symmetric fluid content as
\begin{align}\label{j35}
m_T=&\int^{2\pi}_0 \int^\pi_0 \int^r_0 r^2
e^\frac{\lambda+\nu}{2}\sin
\theta(T^{0(M)}_{0}-T^{1(M)}_{1}-2T^{2(M)}_{2})d\tilde{r}d\theta
d\phi,
\end{align}
where $T^{0(M)}_{0},~T^{1(M)}_{1}$ and $2T^{2(M)}_{2}$ simply show
the components of $T_{\nu}^{\pi(M)}$. The evaluation of these
components along with Eq. \eqref{j33} produce
\begin{align}\label{j36}
&m_T=\bigg[\frac{4 \pi P_r r^3}{F}+m+\frac{1}{32\pi F}\bigg(B+Q
\gamma_7+B_1 \gamma_2+B_2 \gamma_8+B_3 \gamma_9+B_4 \gamma_5+B_5
\gamma_6\bigg)\bigg]e^\frac{\nu+\lambda}{2}.
\end{align}
One can evaluate the Tolman mass in terms of Conformal scalar as
well using Eq. \eqref{j33}, we get
\begin{align}\nonumber
&m_T=e^\frac{\nu+\lambda}{2}\bigg\{\frac{4 \pi
r^3P_r}{F}+\frac{\varepsilon r^3}{3}+\frac{4 \pi
r^3(\rho-P_\perp+P_r)}{3 F} +\frac{r^3}{6
F}\bigg[B+Q(\gamma_1+\gamma_7 -\gamma_9)+B_1 \gamma_2+B_2(2
\gamma_3-\gamma_{10})
\\\label{j37}&+B_3(\gamma_4+\gamma_8-\gamma_{11})+B_4 \gamma_5+B_5
\gamma_6\bigg] +\frac{1}{32 \pi F}\bigg(B+Q\gamma_7+B_1 \gamma_2+B_2
\gamma_3+B_3 \gamma_8 +B_4 \gamma_5+B_5 \gamma_6\bigg)\bigg\}.
\end{align}
Equation \eqref{j37} shows the relation of Tolman mass with the Weyl
scalar in the influence of ADF. Subsequently, we relate Weyl scalar
to the variables belonging to the matter and metric along with extra
degrees of freedom that originate from DHOST theory.

Equation \eqref{j37} relates the Tolman mass to the Weyl scalar
which is of great importance in the field of cosmology. The Riemann
curvature tensor can be divided into two parts, the Ricci tensor and
the Weyl curvature tensor. Within the Weyl curvature tensor, there
is a distinction made between its electric part and magnetic part
which are known as gravitoelectric and gravitomagnetic fields,
respectively. The presence of these fields allows for long-range
interactions in gravitational forces such as tidal forces and
gravitational waves. The concept of a gravitoelectric field is
essentially a relativistic version of Newtonian tidal forces
\cite{danehkar2009significance}. Some researchers have suggested
that certain functions or aspects of the Weyl tensor could provide
an arrow of time in gravity, based on their ability to make
gravitating fluids more inhomogeneous over time. Also, the structure
scalars which are the main focus of our manuscript are evaluated in
terms of Weyl scalar so the calculation of Eq. \eqref{j37} will
provide a path to depict the Tolman mass in terms of the structure
scalars, especially the complexity factor.

Tolman mass \cite{tolman1930weight} is used to describe the ``active
gravitational mass" as it was designed to determine how much energy
has been accumulated in the system. An anisotropic fluid
distribution analysis can be accomplished using its mass formula. It
is an appropriate version of the mass formula for carrying out the
previously stated analysis.

The Misner-Sharp and Tolman's proposed definition of masses, both
yield identical results at the boundary of a star. However, they
differ in terms of their energy content within the interior of the
sphere, unless the distribution is homogeneous and isotropic. The
Misner-Sharp mass has been widely utilized in calculations related
to relativistic stellar collapse due to its usefulness for
computational purposes. On the other hand, Tolman's proposed
definition can be seen as a measure of gravitational mass.

\section{Models}

In this section, we will discuss two models to attain physical
variables within the framework of DHOST theories.

\subsection{Case of $Y_{TF} = 0$}

One model pertains to the vanishing of the scalar function $Y_{TF}$ and the other
focuses on the absence of conformal flatness. In both models, we
will also impose an additional constraint by nullifying radial
pressure to further accomplish solutions.
%\begin{center}
%    $Y_{TF}=0.$
%\end{center}
Primarily, we will explore a spherically symmetric model that
follows the diminishing complexity factor criteria in the framework
of DHOST theories. It would be interesting to explore a stellar
model that deviates from the isotropic and homogeneous solution,
i.e., with the condition $Y_{TF}=0$, as the scalar function $Y_{TF}$
is designed to assess the complexity factor of self-gravitational
fluids. To obtain this specific model in the framework of a higher
order scalar tensor theories, we must introduce an additional
constraint on the system because there could be countless possible
solutions. In this regard, we impose an additional condition,
$P_r=0$, besides ensuring that $Y_{TF}$ is equal to zero. Equation
\eqref{j10}, under the constraint $P_r=0$, turns out to be
\begin{equation}\label{j40}
\nu'=-\frac{re^{\lambda}D}{2F}-\frac{1+e^\lambda}{r},
\end{equation}
where $D=B+Q\gamma_{7}+B_{1}
\gamma_{2}+B_{2}\gamma_{8}+B_{3}\gamma_{9}+B_{4}\gamma_{5}+B_{5}
\gamma_{6}$, is the sum of coupling functions and higher order
scalar terms. The coupling function $D$ depends on the scalar field
$\phi$. Equation \eqref{j40}, when assuming
$\frac{1}{e^{\lambda}}+1=2g(r)$, becomes
\begin{equation}\label{j41}
\nu'=-\frac{rD}{2(2g-1)F}-\frac{2g}{r(2g-1)}.
\end{equation}
The use of Eqs. \eqref{j26} and \eqref{j41} along with the condition
$Y_{TF}=0$, give
\begin{align}\nonumber
&g'\bigg(4r-4rg-\frac{r^3 D}{F}\bigg)+g\bigg(20g-16-16 r^2 A-\frac{2
r^3 D F'}{F^2}+\frac{2 r^3 D'}{F}+\frac{6 r^2 D}{F}\bigg)+ 8 r^2
A-\frac{r^4 D^2}{4 F^2}\\\label{j42}&-\frac{4 r^2 D}{F}-\frac{r^3
D'}{F} +\frac{r^3 D F'}{F^2}=0.
\end{align}
From Eq. \eqref{j42}, the value of $g'$ is
\begin{align}\nonumber
&g'=\frac{-1}{\bigg(4r-4rg-\frac{r^3
D}{F}\bigg)}\bigg[g\bigg(20g-16-16 r^2 A-\frac{2 r^3 D
F'}{F^2}+\frac{2 r^3 D'}{F}+\frac{6 r^2 D}{F}\bigg)+ 8 r^2
A-\frac{r^4 D^2}{4 F^2}\\\label{j43}&-\frac{4 r^2 D}{F}-\frac{r^3
D'}{F} +\frac{r^3 D F'}{F^2}\bigg].
\end{align}
In the presence of modified correction terms, the density and the
tangential pressure for this particular model is calculated as
\begin{align}\nonumber
&\frac{\rho}{F}=\frac{g}{4 \pi r^2}+\frac{1}{(r^2 D-4F+4F
g)}\bigg[g\bigg(\frac{5F g}{\pi r^2}-\frac{2F}{\pi r^2}-\frac{4 A F
}{\pi}+\frac{3D}{2 \pi}+\frac{rD'}{2 \pi}-\frac{rDF'}{2 \pi
F}\bigg)-\frac{D}{\pi}-\frac{r^2 D^2}{16 \pi F}\\\nonumber&+\frac{2
A F}{\pi}-\frac{r D'}{4 \pi}+\frac{r D F'}{4 \pi
F}\bigg]-\frac{D}{8\pi F},
\\\nonumber&\frac{P_{\perp}}{F}=\frac{g}{(-1+2 g)(r^2 D+8F[-1+g])}\bigg[g\bigg(\frac{6F
g}{\pi r^2}-\frac{3F}{\pi r^2}-\frac{2AF}{\pi}+\frac{rD'}{4
\pi}-\frac{rDF'}{4 \pi F}\bigg)-\frac{D}{2 \pi}-\frac{r^2 D^2}{32
\pi F}\\\label{j44}&+\frac{AF}{\pi}+\frac{7D}{8 \pi}-\frac{rD'}{8
\pi}+\frac{rDF'}{8 \pi F}\bigg]-\frac{D}{16\pi F}.
\end{align}
{The condition $Y_{TF}=0$ is a theoretical concept in physics that
provides insights into the stability and structural properties of
spherically symmetric configurations, extending beyond just
cosmological applications. This condition suggests a balance between
energy density and pressure anisotropy, enabling the development of
well-behaved stellar models. Applying vanishing of the scalar
function $Y_{TF}$ to anisotropic models deepens our understanding of
how matter behaves under extreme conditions, like those found in
neutron stars or other compact objects. Contreras \emph{et al.}
\cite{contreras2022uncharged} extended the null complexity condition
to charged anisotropic models, indicating its adaptability across
various matter distributions and gravitational theories.}

{Investigations have explored satisfying the null condition in actual
spherical configurations, identifying conditions for zero complexity
and potential practical applications beyond theoretical constructs.
Polynomial complexity factors have been shown to result in stable
stellar configurations while upholding the null condition,
illustrating how complexity affects stability and feasibility.
Furthermore, the null complexity condition framework offers a
valuable tool for analyzing the interactions between different
matter components under the influence of gravity, which is crucial
for understanding phenomena such as the dynamics of self-gravitating
systems and black hole formation.}

{Applying null complexity models has provided valuable insights
across diverse real-world domains. In social sciences, these
simplified models have facilitated analysis of intricate social
networks, unveiling underlying trends that may not emerge through
conventional methods. Likewise, epidemiological studies employing
null complexity premises have enhanced our understanding of disease
spread by concentrating on pivotal interactions, leading to
streamlined projections of infection rates and outbreak dynamics.}

\subsection{Case of $\varepsilon=0$}

Now, we will utilize the scenario in which the Weyl scalar becomes
zero and derive a solution. The occurrence of a vanishing Weyl
scalar indicates that spacetime conforms to a flat structure. In
other terms, we can affirm that there is a conformal neighborhood
around each point that resembles an open subset of Minkowski
spacetime. Within this framework, to derive a solution, we consider
the Weyl scalar, mentioned in Eq. \eqref{j15}, to be equal to zero.
Consequently, Eq. \eqref{j15} transforms into
\begin{equation}\label{j45}
e^{\lambda-\nu}\bigg[\frac{e^{\nu}\nu'
}{2r}\bigg]^{\prime}=-\bigg[\frac{e^{-\lambda}\nu'
}{2r}\bigg]^{\prime}-\bigg[\frac{1+e^{-\lambda}}{r^2}\bigg]^{\prime}.
\end{equation}
Using the expressions $z=e^{-\lambda}$ and
$\frac{\nu'}{2}=\frac{\mathfrak{b'}}{\mathfrak{b}}$, to get
\begin{equation}\label{j46}
\bigg[\mathfrak{b'}-\frac{\mathfrak{b}}{r}\bigg]z'+2\bigg[\mathfrak{b''}
-\frac{\mathfrak{b'}}{r}+\frac{\mathfrak{b}}{r^2}\bigg]z+\frac{2\mathfrak{b}}
{r^2}=0.
\end{equation}
Equation \eqref{j46} upon integration gives
\begin{equation}\label{j47}
z=\left(\int e^{-\int
\mathfrak{a_1}(r)dr}\mathfrak{a_2}(r)dr+\mathfrak{B}\right)e^{-\int
\mathfrak{a_1}(r)dr},
\end{equation}
where
\begin{eqnarray}\nonumber
&&\mathfrak{a_1}(r)=2
\frac{d}{dr}\left[ln\left(\mathfrak{b}'-\frac{\mathfrak{b}}{r}\right)\right],\\\nonumber
&&\mathfrak{a_2}(r)=\frac{-2\mathfrak{b}}
{\left(\mathfrak{b'}-\frac{\mathfrak{b}}{r}\right)r^2},
\\\nonumber&&\mathfrak{B}\rightarrow \quad integration\quad
constant.
\end{eqnarray}
Equation \eqref{j47} in the form of original variable becomes
\begin{equation}\label{j48}
\frac{\nu'}{2}=\frac{1}{r}+e^{\lambda/2}\sqrt{
e^{-\nu}\omega-\frac{1}{r^2}}.
\end{equation}
Next, by using the conditions $P_{r}=0$ and $\varepsilon=0$ along with
the Eq.\eqref{j40}, we get
\begin{align}\nonumber&
9+10 e^\lambda+e^{2 \lambda}+3 r \lambda'-e^\lambda r
\lambda'+\frac{e^{2 \lambda} r^4 D^2}{4 F^2}+\frac{e^\lambda r^2
D}{F}+\frac{e^{2 \lambda} r^2 D}{F}-\frac{e^\lambda r^3
D^{\prime}}{F} \\\label{j49}& +\frac{e^\lambda r^3 D
F^{\prime}}{F^2}-\frac{e^\lambda r^3 D \lambda^{\prime}}{2 F}=0.
\end{align}
Equation \eqref{j49}, with the use of expression
$e^{-\lambda}+1=2g(r)$, becomes
\begin{align}\nonumber&
g'\bigg(8r-12rg+\frac{r^3 D}{F}\bigg)+g\bigg(36 g-16 -\frac{2 r^3
D'}{F}+\frac{2 r^3 DF'}{F^2}+\frac{2 r^2 D}{F}\bigg)\\\nonumber &
+\frac{r^3 D'}{F}-\frac{r^3 D F'}{F^2}+\frac{r^4 D^2}{4 F^2}=0.
\end{align}
The state variables will become
\begin{align}\nonumber
\frac{\rho}{F}&=\frac{1}{[r^2 D+4 F(2-3 g)]}\bigg[g\bigg(\frac{6
F}{\pi r^2}-\frac{12 F g}{\pi r^2}-\frac{D}{4 \pi}+\frac{rD'}{2
\pi}-\frac{r DF'}{2 \pi F}\bigg)-\frac{r^2 D^2}{16 \pi F}-\frac{r
D'}{4 \pi}\\\label{j50}&+\frac{r D F'}{4 \pi F}\bigg]-\frac{D}{8\pi
F}
\\\nonumber \frac{P_{\perp}}{F}&=\frac{1}{[r^2 D+4 F(2-3 g)](-1+2
g)}\bigg[g^2\bigg(\frac{3 F}{\pi r^2}-\frac{6 F g}{\pi r^2}-\frac{D
}{8 \pi}+\frac{r  D'}{4 \pi}-\frac{r DF'}{4 \pi
F}\bigg)\\\label{j51} &+g\bigg(-\frac{r^2 D^2}{32 \pi
F}-\frac{rD'}{8 \pi}+\frac{r DF'}{8 \pi
F}\bigg)\bigg]-\frac{D}{16\pi F}.
\end{align}
Putting the value of $\nu'$ and $e^{\frac{\lambda}{2}}$ in Eq.
\eqref{j48}, we get
\begin{align}\label{j52}
e^{\nu}=\frac{\omega
r^{2}(2g-1)}{\bigg(\frac{r^{2}D}{4F}-1+3g\bigg)^{2}+2g-1}.
\end{align}
The results of these models will reduce to $GR$ result evaluated in
\cite{herrera2021hyperbolically} on substituting $F=1,~D=0$ and
$A=0$.

\subsection{Comparison of Models: $Y_{TF}=0$  v/s $\varepsilon=0$}

{The vanishing complexity factor and the vanishing conformal scalar
are both crucial concepts in the study of complex systems within
theoretical physics, particularly in the context of self-gravitating
systems such as stars and other astrophysical objects. These
concepts provide different perspectives on the structure and
stability of these complex systems.}

{Systems with a vanishing complexity factor tend to exhibit greater
stability due to their uniform energy density and isotropic
pressure. This uniformity reduces the likelihood of instabilities
that can arise from anisotropic distributions or intricate
interactions within the matter, making vanishing complexity factor
models more preferable for modeling self-gravitating systems like
neutron stars. The ability of zero complexity factor models to avoid
singularities enhances their physical viability in representing
real-world astrophysical objects. In contrast, a vanishing conformal
scalar can imply certain geometric symmetries that simplify
equations, but it does not inherently guarantee stability. The
implications depend heavily on the specific context and additional
conditions imposed on the system.}

{While both models provide valuable insights into gravitational
theories, the vanishing complexity factor stands out as a more
effective tool for constructing stable and physically viable models
in astrophysical contexts. Its capacity to simplify complex matter
distributions while ensuring stability makes it a preferred choice
among researchers studying self-gravitating systems. The vanishing
complexity factor offers a more direct and robust approach to
modeling the behavior of these complex systems, leading to more
reliable and physically grounded representations of astrophysical
phenomena.}\\\\

A special case of DHOST theories can also be discussed by
implementing constraints on arbitrary functions. The unknown
functions $F,~B$ and $Q$ can be chosen arbitrarily, while the
remaining ones must satisfy specific degeneracy conditions to ensure
that there is only one scalar degree of freedom. We can consider
these degeneracy conditions which are studied in
\cite{achour2016degenerate}. It was demonstrated that physically
viable theories in the case of quadratic DHOST belong to class NI
\cite{crisostomi2016extended}. Bearing in mind $X=X_{0}$, the
following degeneracy conditions can be taken
\begin{align}\label{j38}
B_{2}=-B_{1}=X_{0}Q,\quad \quad B_{3}=-B_{4}=-2Q, \quad \quad
B_{5}=0.
\end{align}
In our present work, we have considered anisotropic pressure as we
wanted to discuss complexity of the self gravitating objects.
Additionally, we have performed calculations for two models
exhibiting the constraint of conformal flatness and minimal
complexity. The implementation of the conditions mentioned in Eq.
\eqref{j38} will be explored in future research along with the
change in symmetry or supplementary constraints.

\section{Conclusions and Discussions}

In astrophysics, understanding the fundamental structure of tightly
packed items like black holes and neutron stars requires an in-depth
comprehension of anisotropic fluid distributions and their
complexity, while anisotropy is important in these extreme scenarios
because it affects singularity creation, collapse, and the
stability. Comprehending how anisotropic pressures act differently
from isotropic ones under modified gravity which may result in
discernible variations in gravitational wave signatures or energy
emissions from such objects allows us to better understand the
structure scalars in DHOST theories.

In the current study, the gravitational sector is non-minimally
coupled with derivatives of scalar fields. This is motivated by the
coupling of the matter to the Einstein tensor
\cite{asimakis2022modified} as well as by the coupling of the matter
sector to the Ricci scalar. This manuscript mainly has three aims
which are as follows.

To begin with, we did an extensive analysis of the general framework
of DHOST theory. For this purpose, we considered action integral and
varied it concerning the metric tensor. Moreover, using spherically
static anisotropic matter content we have evaluated its
gravitational field and non-conserved equations. We determine the
structure scalars in the framework of DHOST theory. These scalars
appear to be particularly well adapted for the description of
self-gravitating fluids. This set of scalars which are acquired from
the (1+3) decomposition of the Curvature tensor is the major goal of
this article. The mass definitions, such as the Misner-Sharp and
Tolman masses, are crucial for understanding energy contributions in
astrophysical objects which are more dense and cosmological models,
specifically, these masses aid in determining the system's overall
energy content, which encompasses both gravitational and matter
energy. Lastly, by extending these definitions to DHOST theories, we
study how corrections owing to changed gravity affect the
gravitational mass and energy distribution in astronomical objects.
This might directly affect how observational data are interpreted,
such as the dynamics of galaxies or the mass-radius relation in
neutron stars, which may be covered in the future.

It is not easy to define the complexity of multiple (but similar)
systems. This idea has undergone extensive study and application to
several scientific contexts since it is connected to numerous
fascinating features of the configuration of nature
\cite{lopez1995statistical,crutchfield2000comment,sanudo2009complexity,de2012entropy}.
Motivated by the work of  Herrera \cite{herrera2009structure} and
the significance of scalar-tensor theories, we construct the unique
notion of complexity within the context of DHOST theory.  For this
purpose, we implemented the orthogonal decomposition method.  Hence,
we can get a simpler expression for the Riemann curvature tensor and
deduce the structure scalars from it. We would also like to draw
attention to the fact that we have restricted ourselves to a few
sectors of the large family of modified gravity theories. It would
be fascinating to explore other areas of this large family of
modified gravity theories. Along with this, we have discussed two
models in which we deal with conformal flatness and minimal
complexity. For these specific models, we have evaluated the
physical variables in the presence of modified correction terms. The
results of these models will reduce to $GR$ on substituting
$F=1,~D=0$ and $A=0$, discussed in \cite{herrera2021hyperbolically}.

Many fields of modern physics rely greatly on the scalar field.
After Newton's gravity, which possesses a scalar potential field,
its use in cosmology was first noted in Nordstr\"{o}m's
investigations \cite{nordstrom1913trage}. The initial stage of rapid
growth is driven by one or more scalar fields. Scalar fields, which
are akin to early inflationary ideas, have therefore been used in
several research to explain dark energy. One often begins with the
Einstein-Hilbert Lagrangian and extends it in different ways to
develop gravitational modifications. One may take into account ideas
in which the geometric and non-geometric aspects of the action are
linked. The generic scalar-tensor theory and Horndeski/Galileon
theory is the two simplest GR models. They both are non-minimally
linked and non-minimal derivatively coupled scalar fields.

In recent years, there has been growing interest in using modified
gravity theories, such as DHOST theories, to explain the properties
of different compact objects. This present study could be a
significant addition to the expanding literature on DHOST theories
and their applications in astrophysics. Through extensive analysis,
it could not only identify the distinctive characteristics of these
theories but also elucidate their enormous potential for further
exploration. It may also offer an invaluable resource for
researchers pursuing studies related to compact objects' behaviour
under intense gravitational fields. The findings presented herein
have far-reaching implications that could significantly advance our
understanding of the complex nature of celestial bodies.
Furthermore, it may enhance our ability to unravel some of the
universe's most perplexing mysteries.

There could exist a logical progression in our study towards
formulating a methodology for cylindrical, hyperbolic, axisymmetric
or non-static arrangements in future. Despite its complexity,
initial advancements have been accomplished within the DHOST
framework through the implementation of anisotropic spherical
symmetry. Even more, investigations could potentially lead to a
better understanding of the dynamics and behaviour of
self-gravitating systems. It may also contribute to groundbreaking
developments in the field of astrophysical simulations as well as in
related industries. Furthermore, by investigating how DHOST theories
impact large-scale celestial evolution, these discoveries can be
applied to the field of cosmology, while anisotropic stress, for
example, could have an impact on the dynamics of the cosmic web or
structure creation in the early cosmos. Moreover, gravitational wave
propagation in modified gravity theories may disclose GR deviations
that could be observed by Square Kilometer Array or LIGO, among
other present day or future detectors. Application of our findings
in the context of DHOST theories to real-world astrophysical
instances, including black holes, neutron stars, and cosmological
models, could pave new avenues for future investigations.

\section*{Appendix}
The values of $\gamma_{i}'s$ $(i=1-10)$ appeared in
Eq.\eqref{j9}-\eqref{j11} are given as
\begin{align}\nonumber
\gamma_1&=\Box \phi, \\\nonumber
\gamma_2&=e^{-2\lambda}\bigg(\frac{\nu^{\prime} \phi^{\prime
2}}{4}+\phi^{\prime \prime}+\frac{\lambda^{\prime} \phi^{\prime
2}}{4}-\phi^{\prime \prime} \lambda^{\prime} \phi^{\prime}+\frac{2
\phi^{\prime^ {2}}}{r^{2}}\bigg), \\\nonumber\gamma_3&=(\Box
\phi)^{2},
\\\nonumber \gamma_4&=e^{-2\nu-\lambda} \phi^{\prime
2}\bigg(\phi^{\prime \prime}-\frac{\lambda^{\prime}
\phi^{\prime}}{2}\bigg)\Box \phi,\\\nonumber \gamma_5&=e^{-3\nu}
\phi^{\prime^{2}}\bigg(\phi^{\prime \prime 2}+\frac{\lambda^{\prime
2} \phi^{\prime 2}}{4}-\phi^{\prime \prime} \lambda^{\prime}
\phi^{\prime}\bigg),\\\nonumber \gamma_6&=\bigg(e^{-2\nu}
\phi^{\prime
2}\bigg)^2\bigg(\phi^{\prime\prime}+\frac{\lambda^{\prime 2}
\phi^{\prime^{2}}}{4}-\phi^{\prime \prime} \lambda^{\prime}
\phi^{\prime}\bigg),\\\nonumber \gamma_7&=\Box
\phi+2\phi''e^{-\lambda},\\\nonumber \gamma_8&=(\Box
\phi)^{2}+4(\Box \phi)\phi''e^{-\lambda},\\\nonumber
\gamma_9&=e^{-2\nu}\phi'^{2}\bigg(\phi''-\frac{\lambda'\phi'}{2}\bigg)[\Box
\phi+\phi''],\\\nonumber \gamma_{10}&=(\Box \phi)
e^{-2\nu}\phi'^{2}\bigg(\phi''-\frac{\lambda'\phi'}{2}\bigg).
\end{align}
The value of $I_{1}$ comes in Eq.\eqref{j19} and \eqref{j20} is
evaluated as
\begin{align}\nonumber
I_{1}&=B+Q\Box \phi+B_{1}(\nabla_{\alpha} \nabla_{\beta}\phi)
(\nabla^{\alpha} \nabla^{\beta} \phi)+B_2(\Box \phi)^2+B_{3}(\Box
\phi) \phi^{\rho} \phi^{\lambda}(\nabla_{\rho} \nabla_{\lambda}
\phi)+B_{4}
\\\nonumber&\times\left(\phi^{\lambda}\phi_{\lambda\rho}\phi^{\rho\gamma}
\phi_{\gamma})+B_{5}[(\nabla_{\rho}\nabla_{\lambda}\phi)(\phi^{\rho}
\phi^{\lambda})\right]^{2}.
\end{align}
The value of $I_2$ occurred in Eq.\eqref{j26} is defined as
\begin{align}\nonumber
&I_{2}=\frac{h_{\alpha\beta}}{2 F}I_{1}+\bigg[h_{\alpha}^{\pi}
h_{\beta}^{\nu}(\partial_{\pi} \partial_{\nu}\phi)+g_{\alpha \beta}
V_{\gamma} V^{\gamma}(\partial^{\gamma} \partial_{\delta}
\phi)\bigg]\times\bigg[-Q-2B_{2}\Box \phi-B_{3}\phi^{\alpha}
\phi^{\beta} \phi_{\alpha\beta}\bigg] -\frac{h^{\pi}_{\alpha}
h^{\nu}_{\beta}}{2F}I_{1}+h_{\alpha}^{\pi}
h_{\beta}^{\nu}\\\nonumber&\times\bigg[\Box \phi-(\partial_\pi
\partial_\delta \phi) V^\pi V^\delta-V^\nu
V_\gamma(\partial^{\gamma} \partial_{\nu} \phi)+4 V_{\gamma}
V^{\delta}(\partial^{\gamma} \partial_{\delta} \phi)\bigg]
\bigg[-Q-2B_{2}\Box
\phi-B_{3}\phi^{\alpha}\phi^{\beta}\phi_{\alpha\beta}\bigg].
\end{align}
The value of $I_3$ occurred in Eq.\eqref{j28} is calculated as
\begin{align}\nonumber
& I_3=-h_{\xi}^{\pi}h_{\eta}^{\nu}\bigg[Q+2B_{2}\Box
\phi+B_{3}\phi^{\alpha} \phi^{\beta} \phi_{\alpha
\beta}\bigg]\bigg[\epsilon_{\pi}^{\beta\delta} \epsilon_{\alpha
\delta \nu}(\partial^{\alpha}
\partial_{\beta}\phi)-\epsilon_{\pi}^{\beta\delta}\epsilon_{\alpha\beta\nu}(\partial^{\alpha}
\partial_{\delta} \phi)-\epsilon_{\pi}^{\beta \delta}
\epsilon_{\delta \gamma \nu}(\partial^{\gamma} \partial_{\beta}
\phi)\\\nonumber &+\epsilon_{\pi}^{\beta \delta} \epsilon_{\beta
\gamma \nu}(\partial^{\gamma} \partial_{\delta}
\phi)\bigg]-\frac{h_\xi\eta}{6F}\bigg[Q+2B_{2}\Box
\phi+B_{3}\phi^{\alpha}\phi^{\beta}\phi_{\alpha
\beta}\bigg]\bigg[h_{\alpha}^{\beta}(\partial_{\beta}\partial^{\alpha}\phi)+h_{\gamma}^{\beta}
(\partial_{\beta}\partial^{\gamma}\phi)+h_{\alpha}^{\delta}
(\partial_{\delta}
\partial^{\alpha}\phi)\\\nonumber&+h_{\gamma}^{\delta}(\partial_{\delta}
\partial^{\gamma}\phi)\bigg].
\end{align}
\section*{Acknowledgement}
The work of KB was partially supported by the JSPS KAKENHI Grant
Number 21K03547, 23KF0008 and 24KF0100.

\section*{Data Availability Statement}
All data generated or analyzed during this study are included in this published article.

\end{document}